\documentclass[preprint,preprintnumbers,amsmath,amssymb]{revtex4}
\usepackage{graphicx}
\usepackage{epstopdf}
\usepackage{dcolumn}
\usepackage{bm}
\usepackage{color}
 \usepackage{array,multirow,graphicx}

\begin{document}

\title{Electronic structure and magnetic properties of the graphene/Ni$_3$Mn/Ni(111) trilayer}

\author{Elena Voloshina,$^{1,2,}$\footnote{Corresponding author. E-mail: voloshina@shu.edu.cn} Qilin Guo,$^{1,2}$ Beate Paulus,$^2$ Stefan B\"ottcher,$^3$ Karsten Horn,$^3$ Changbao Zhao,$^{4,5}$ Yi Cui,$^{4,}$\footnote{Corresponding author. E-mail: ycui2015@sinano.ac.cn} and Yuriy Dedkov$^{1,}$\footnote{Corresponding author. E-mail: dedkov@shu.edu.cn}}

\affiliation{$^1$\mbox{Department of Physics, Shanghai University, 99 Shangda Road, 200444 Shanghai, China}}
\affiliation{$^2$\mbox{Freie Universit\"at Berlin, Physikalische und Theoretische Chemie, 14195 Berlin, Germany}}
\affiliation{$^3$\mbox{Fritz Haber Institut of the Max Planck Society, 14195 Berlin, Germany}}
\affiliation{$^4${Vacuum Interconnected Nanotech Workstation, Suzhou Institute of Nano-Tech and Nano-Bionics, Chinese Academy of Sciences, 215123 Suzhou, China}}
\affiliation{$^5${Nano Science and Technology Institute, University of Science and Technology of China, 215123 Suzhou, China}}

\date{\today}

\begin{abstract}

Experimental and theoretical studies of manganese deposition on graphene/Ni(111) shows that a thin ferromagnetic Ni$_3$Mn layer, which is protected by the graphene overlayer, is formed upon Mn intercalation. The electronic bands of graphene are affected by Ni$_3$Mn interlayer formation through a slight reduction of $n$-type doping compared to graphene/Ni(111) and a suppression of the interface states characteristic of graphene/Ni(111). Our DFT-based theoretical analysis of interface geometric, electronic, and magnetic structure gives strong support to our interpretation of the experimental scanning tunneling microscopy, low energy electron diffraction, and photoemission results, and shows that the magnetic properties of graphene on Ni(111) are strongly influenced by Ni$_3$Mn formation.\\

This document is the unedited author's version of a Submitted Work that was subsequently accepted for publication in J.  Phys. Chem. C., copyright $\copyright$ American Chemical Society after peer review. To access the final edited and published work see doi: 10.1021/acs.jpcc.9b00942.

\end{abstract}

\maketitle

\section{Introduction}

Recently, graphene-metal interfaces have attracted a lot of attention~\cite{Batzill:2012,Dedkov:2015kp} because many promising ideas of technological applications of these systems have appeared in the literature. For example, the synthesis of large layers of high-quality graphene (gr) on metals~\cite{Bae:2010,Liu:2011bl,Lee:2017cm} is considered as the most promising method, which in future may be implemented in different areas of micro-electronics and nano-technology~\cite{Kim:2009a,Ryu:2014fo}. Also, graphene is found to be one of the best one-atom-thick protection layers for different substrates, metallic as well as semiconducting~\cite{Dedkov:2008d,Dedkov:2008e,Sutter:2010bx,Weatherup:2015cx,Campbell:2018ii,Tesch:2018hm,Zhou:2018ji}.

Considering the electronic structure of graphene-metal systems, graphene on these substrates always becomes either $n$ or $p$ doped, demonstrating, in some cases, also the overlap of the graphene $\pi$ states and valence band states of the underlying material (\mbox{space-,} \mbox{energy-,} and \mbox{$k$-vector-overlapping}). 
Such cases are described as \textit{strongly} 
interacting graphene with metals~\cite{Voloshina:2012c,Voloshina:2014jl}. 
In graphene/metal system belonging to this group, the electronic structure of graphene is heavily disturbed due to the above mentioned overlap of the valence band states of graphene and metal, leading to the loss of the free-standing character of the electronic structure in graphene.

Among the \textit{strongly} interacting systems, the graphene-ferromagnet (gr-FM) interfaces are the most exciting due to their perspectives for spintronic applications. For example, it was predicted that a stack of graphene layers can be used as an effective spin filter in FM-gr$_n$-FM sandwiches~\cite{Karpan:2007,Karpan:2008}. The overlap of the valence band states of graphene and the FM material drastically reduces spin-filtering effect for the single graphene layer, which however can be improved via placing layers of noble metals or h-BN between graphene and the FM~\cite{Karpan:2008,Karpan:2011kv}. However, the overlap of the electronic states at the interface between graphene and FM materials (Ni, Co) allows to observe the induced spin-polarization of the graphene $\pi$ states that was confirmed in a series of spectroscopic experiments for gr/Ni(111) and gr/Co(0001)~\cite{Weser:2010,Dedkov:2010jh,Weser:2011,Marchenko:2015ka,Usachov:2015kr} and this effect can be used for an effective spin injection and manipulation in graphene.

The electronic structure of the gr-FM interface can be tailored via intercalation of different species with the aim to prepare different graphene-based heterostructures. Here, e.\,g., the intercalation of Fe leads to an increase of the induced magnetic moment in graphene~\cite{Weser:2011}, the intercalation of $sp$-metals decouples graphene from the FM with the controllable modification of the graphene band structure around the Dirac point~\cite{Dedkov:2001,Varykhalov:2008,Varykhalov:2010a,Voloshina:2011NJP}. The intercalation of oxygen leads to the formation of thin layers of oxides and the resulting epitaxial systems might be used in future spintronics applications~\cite{Omiciuolo:2014dn,Dedkov:2017jn}. Moreover, the insertion of other magnetic atoms with open $d$-shells into the gr-FM interface might lead to the formation of ordered gr/FM-alloy systems with interesting properties.

Here we present systematic electronic structure studies of the system formed after intercalation of Mn in gr/Ni(111). It is found that an ordered $(2\times2)$ structure with respect to gr/Ni is formed at the interface, indicating the formation of a thin layer of the strong FM Ni$_3$Mn alloy. This alloy is effectively protected by a graphene monolayer and the electronic structure of this system was studied via application of different spectroscopic methods, accompanied by theoretical calculations. Very good agreement is found between theory and experiment.

\section{Results and discussion}

\begin{figure}[t]
	\includegraphics[width=\columnwidth]{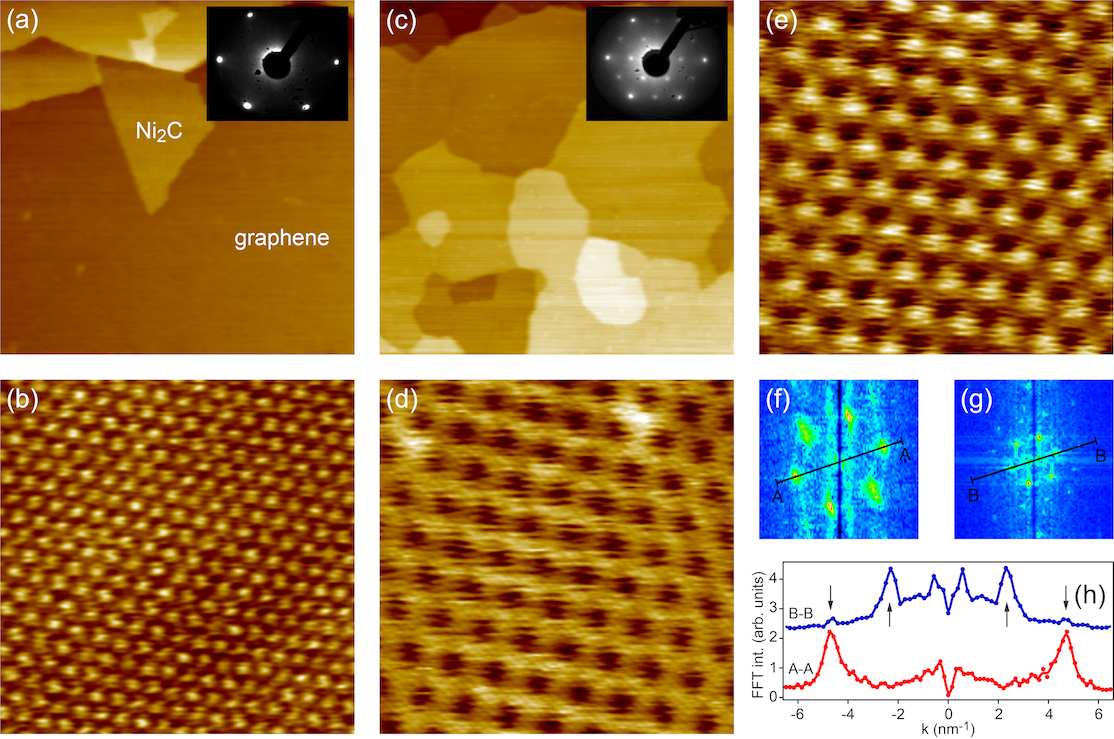}
	\caption{(a,b) STM images of gr/Ni(111); scanning parameters: (a) $150\mathrm{nm}\times150\mathrm{nm}$, $U_T=+0.1$\,V, $I_T=0.31$\,nA, (b) $4\mathrm{nm}\times4\mathrm{nm}$, $U_T=+0.005$\,V, $I_T=0.52$\,nA. (c-e) STM images of the system obtained after intercalation of Mn in gr/Ni(111); scanning parameters: (c) $135\mathrm{nm}\times135\mathrm{nm}$, $U_T=+1.25$\,V, $I_T=0.1$\,nA, (d) $4\mathrm{nm}\times4\mathrm{nm}$, $U_T=+0.022$\,V, $I_T=0.11$\,nA, (e) $4\mathrm{nm}\times4\mathrm{nm}$, $U_T=+0.01$\,V, $I_T=0.31$\,nA. Insets of (a) and (c) show the respective LEED images collected at $80$\,eV and $110$\,eV. (f,g) FFT images of (b,e), respectively. (h) FFT intensity profiles taken in (f) and (g) with down- and up-arrows marking $(1\times1)$ and $(2\times2)$ periodicities, respectively.}
	\label{grNi_grMnNi_STM_LEED}
\end{figure}

\textit{LEED and STM}. The intercalation of Mn atoms in gr/Ni(111) and formation of the intercalation-like systems was studied in a series of experiments using STM, LEED and XPS. Fig.~\ref{grNi_grMnNi_STM_LEED} shows large and small scale STM images of gr/Ni(111) (a,b) and of the system obtained after annealing of Mn/gr/Ni(111) at $400^\circ$\,C (c-e). The respective LEED images of the studied systems are presented as insets for large scale STM data. Due to the lattice match of graphene and Ni(111) (the difference is approximately $1$\%), LEED and STM of the gr/Ni(111) system demonstrate $(1\times1)$ in-plane periodicities indicating also the high quality of this system on the large and the atomic scale. The number of defects in the graphene layer for this system is very small; however, some areas associated with the presence of the Ni$_2$C phase (small scale STM images of this area are not shown here) can be detected on the surface; this is also supported by the XPS data and consistent with the previously published results for the same system~\cite{Larciprete:2016gf,Bignardi:2017aa,Dedkov:2017jn}. The LEED and STM images of the system obtained after intercalation of Mn in gr/Ni(111) demonstrate $(2\times2)$ periodicities with respect to the ones of the parent system (Fig.~\ref{grNi_grMnNi_STM_LEED}). Small scale STM images presented in (d) and (e) were collected at slightly different bias voltages that might cause the change in the imaging contrast. Fast Fourier transformation (FFT) images of the atomically resolved STM data for gr/Ni(111) before and after intercalation of Mn are shown in panels (f) and (g), respectively, with the extracted corresponding intensity profiles shown in (h), and they also confirm the formation of a $(2\times2)$ structure for the gr-Mn-Ni(111) interface.

\begin{figure}[t]
	\includegraphics[width=0.9\columnwidth]{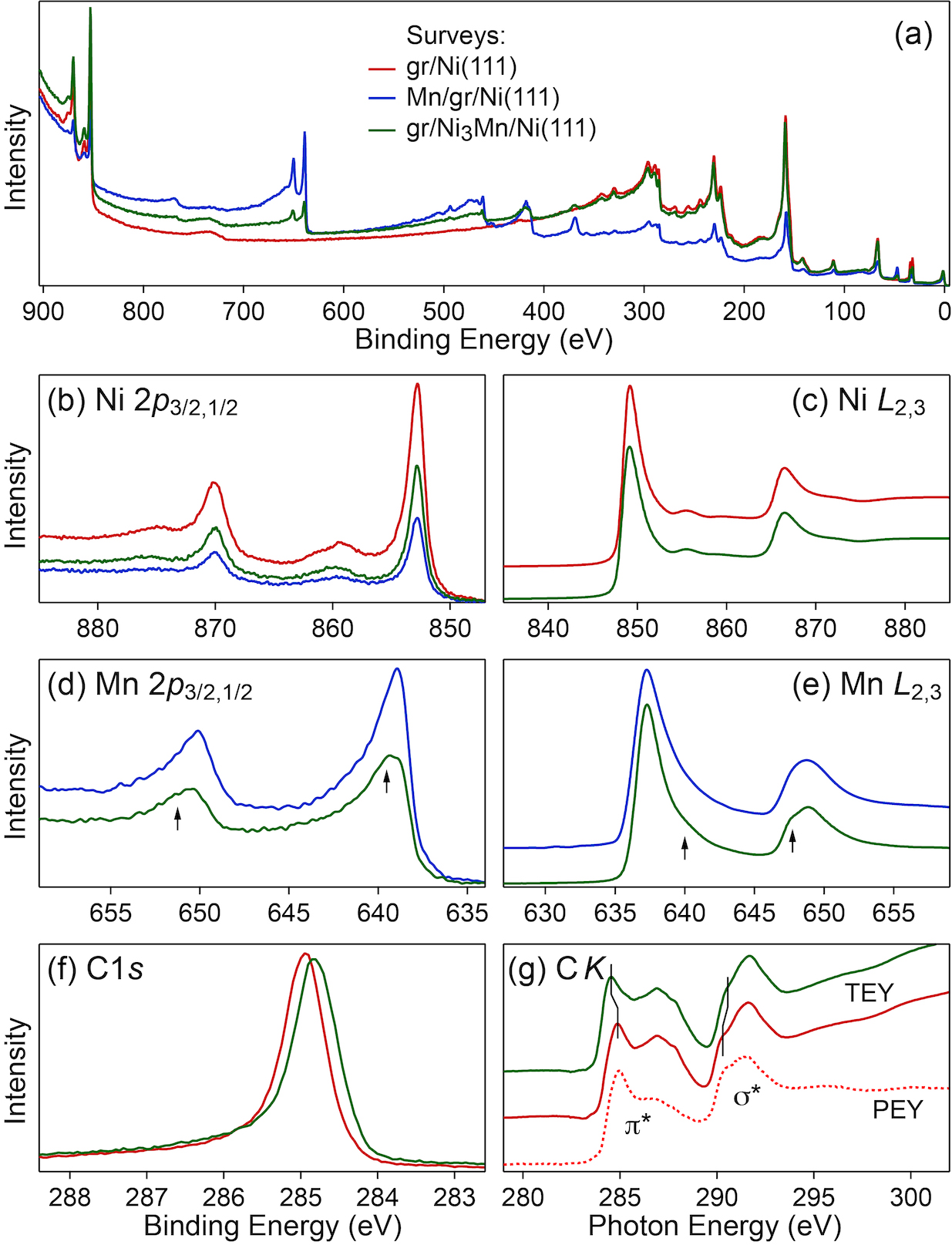}
	\caption{XPS and NEXAFS spectra for systems under study: (a) XPS surveys, (b) XPS Ni\,$2p_{3/2,1/2}$, (c) NEXAFS Ni\,$L_{2,3}$, (d) XPS Mn\,$2p_{3/2,1/2}$, (e) NEXAFS Mn\,$L_{2,3}$, (f) XPS C\,$1s$, and (g) NEXAFS C\,$K$. Red, blue, and green colours shows spectra for gr/Ni(111), Mn/gr/Ni(111), and gr/Ni$_3$Mn/Ni(111), respectively.}
	\label{grNi_grMnNi_XPS_NEXAFS_summary}
\end{figure}

\textit{XPS and NEXAFS}. Figure~\ref{grNi_grMnNi_XPS_NEXAFS_summary} compiles the results of the XPS studies: (a) survey spectra before and after Mn intercalation in gr/Ni(111) and the respective core-level spectra, (b) Ni\,$2p$, (d) Mn\,$2p$, and (f) C\,$1s$. The corresponding NEXAFS spectra are shown on the right-hand side: (c) Ni\,$L_{2,3}$, (e) Mn\,$L_{2,3}$, and (g) C\,$K$. The process of Mn intercalation is characterized by the corresponding intensity variations of XPS lines as well as line shape modifications for the C\,$1s$ and Mn\,$2p$ lines. 

For the gr/Ni(111) system the binding energy of the C\,$1s$ line is $284.95\pm0.05$\,eV, which is in good agreement with previously measured values~\cite{Weser:2010}. The higher binding energy for this line compared to the neutral graphene ($284.1$\,eV~\cite{Schroder:2016eb}) indicates the strong $n$ doping of graphene in the former case. After intercalation of Mn in gr/Ni(1111) the intensity of the C\,$1s$ line is almost fully restored and a small energy shift of this line of $\approx140$\,meV to smaller binding energies is detected indicating a small reduction of $n$-doping of the graphene layer.

Intercalation of Mn in gr/Ni(111) leads to a slight broadening of both Mn spin-orbit split peaks compared to the Mn\,$2p$ XPS spectra for the non-intercalated case. This increase of the width of the Mn\,$2p_{3/2,1/2}$ emission lines as well as the previously observed $(2\times2)$ structures in LEED and STM data can be assigned to the formation of a thin layer of Ni$_3$Mn alloy at the gr/Mn-Ni(111) interface~\cite{Durr:1997fa,Allen:2001,Rader:2001di,Manju:2010cs}. The formation of the NiMn phase can be ruled out as it would show $p(2\times1)$ patterns for the single domain and $(2\times2)$ patterns from all possible three fold rotated domains; however, a $p(2\times1)$ structure was not observed in the STM experiments, which reflect the local (domain) structure (see Fig.~\ref{grNi_grMnNi_STM_LEED}). In the XPS spectra one can assign the broadening of the Mn\,$2p$ lines to the appearance of emission satellites (marked by arrows in (d)). We believe that the increased Mn-Mn distance in the Ni$_3$Mn(111) plane compared to bulk Mn leads to an increase of the Mn $3d-3d$ correlations and a slightly reduced screening for the core hole as seen in other systems~\cite{Durr:1997fa,Rader:2001di,Manju:2010cs}. The same is valid for the description of the Mn\,$L_{2,3}$ NEXAFS spectra where additional weak correlation satellites are visible at $\approx640$\,eV and $\approx647.5$\,eV (marked by arrows) (see Fig.~\ref{grNi_grMnNi_XPS_NEXAFS_summary}(e) and Refs.~\citenum{Durr:1997fa,Manju:2010cs}).

The C\,$K$ NEXAFS spectra obtained for gr/Ni(111) measured in the PEY and TEY modes (Fig.~\ref{grNi_grMnNi_XPS_NEXAFS_summary}(g), red dashed and solid curves, respectively) are in very good agreement with previously published data~\cite{Weser:2010,Voloshina:2013cw,Matsumoto:2013eu,Verbitskiy:2015kq}. Two structures in the ranges $284-288$\,eV and $289.5-294.5$\,eV are assigned to the $1s\rightarrow\pi^*$ and $1s\rightarrow\sigma^*$ transitions, respectively. A closeup analysis of the NEXAFS spectra of gr/Ni(111) and gr/Ni$_3$Mn/Ni(111) shows that the $1s\rightarrow\pi^*$ absorption edge for the latter system is shifted to smaller photon energies by $\approx300$\,meV and the energy separation between the peaks of $\pi^*$ and $\sigma^*$ features is increased by $\approx430$\,meV (the respective peak shifts are marked in (g)). This effect is assigned to the reduced $n$-doping of a graphene layer and its slightly smaller buckling as previously discussed~\cite{Voloshina:2013cw,Voloshina:2011NJP}.

\begin{figure}[t]
	\includegraphics[width=\columnwidth]{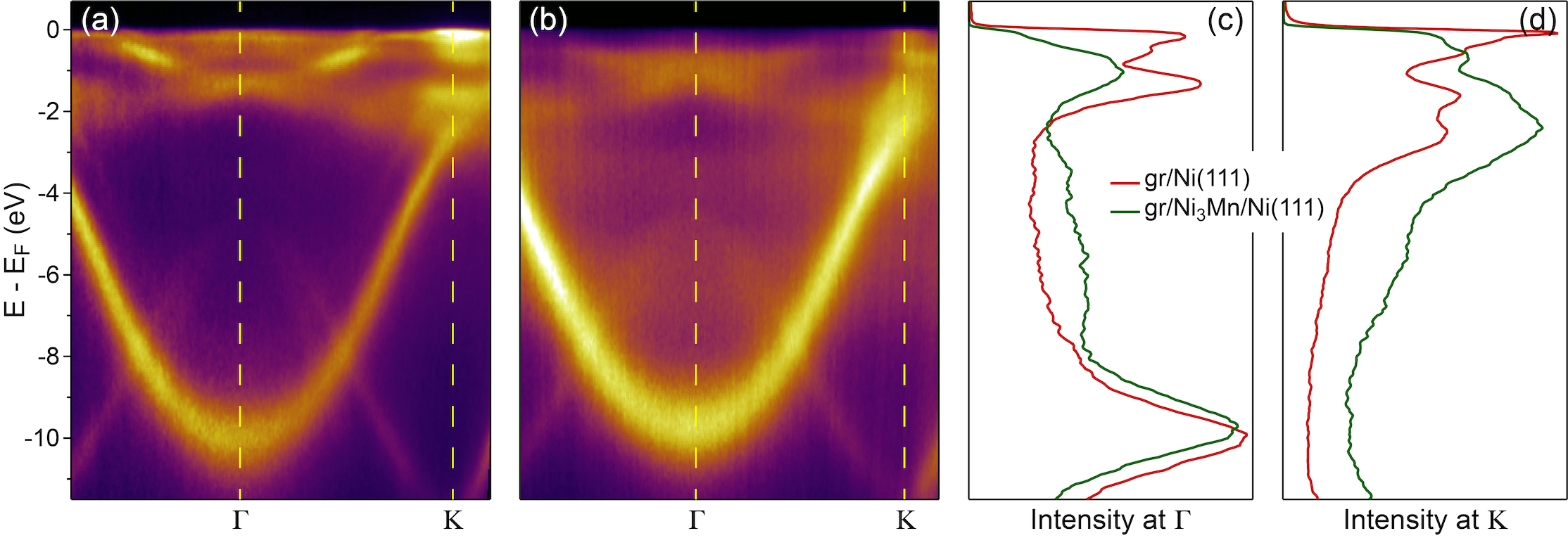}
	\caption{ARPES intensity maps along $\Gamma-\mathrm{K}$ for (a) gr/Ni(111) and (b) gr/Ni$_3$Mn/Ni(111). Photon energy used in the experiment is $h\nu=63$\,eV. Intensity profiles extracted at the $\Gamma$ and $\mathrm{K}$ points for both systems are shown in (c) and (d), respectively.}
	\label{ARPES_full}
\end{figure}

\textit{ARPES}. Figure~\ref{ARPES_full} shows ARPES intensity maps measured for (a) gr/Ni(111) and (b) gr/Ni$_3$Mn/Ni(111) at a photon energy of $h\nu=63$\,eV. The corresponding intensity profiles taken at the $\Gamma$ and $\mathrm{K}$ points of the graphene-derived BZ are presented in panels (c) and (d), respectively. Compared to free-standing graphene, the electronic structure of graphene on Ni(111) is strongly modified due to the $n$-doping and a strong spatial and energy overlap of the graphene $\pi$ and Ni $3d$ valence band states. As a result, the states of graphene are shifted to higher binding energies and the bottom of the $\pi$ band is found at $E-E_F=-9.98\pm0.01$\,eV at the $\Gamma$ point, consistent with previously published data~\cite{Dedkov:2010jh,Voloshina:2011NJP,Verbitskiy:2015kq,Dedkov:2017jn} (Fig.~\ref{ARPES_full}(c)). The overlap and hybridization of the valence band states of graphene and Ni leads to the formation of the so-called interface states below and above $E_F$~\cite{Bertoni:2004,Voloshina:2011NJP}; they can be clearly recognised as a series of peaks in the energy range $E-E_F=0\,... -2.2$\,eV in the intensity profile taken at the $\mathrm{K}$ point for gr/Ni(111) (Fig.~\ref{ARPES_full}(d)). The partial charge transfer from ferromagnetic Ni to graphene and hybridization between valence band states leads to the appearance of the magnetic moment of C atoms in graphene as confirmed in the experiment~\cite{Weser:2010,Weser:2011,Matsumoto:2013eu}.

The intercalation of Mn in gr/Ni(111) followed by the formation of a thin layer of Ni$_3$Mn leads to the shift of the graphene $\pi$ band to smaller binding energy, indicating the slightly smaller $n$-doping of graphene compared to the parent system. The bottom of the $\pi$ band at the $\Gamma$ point in this case is located at $E-E_F=-9.71\pm0.01$\,eV, i.\,e. it is shifted by   $0.27\pm0.02$\,eV to smaller binding energies (Fig.~\ref{ARPES_full}(c)). The electronic structure around the $\mathrm{K}$ point is also modified as manifested by the suppression of the strong intensity associated with the interface states found in the previous system (Fig.~\ref{ARPES_full}(d)). 

\begin{figure}[t]
	\includegraphics[width=0.6\columnwidth]{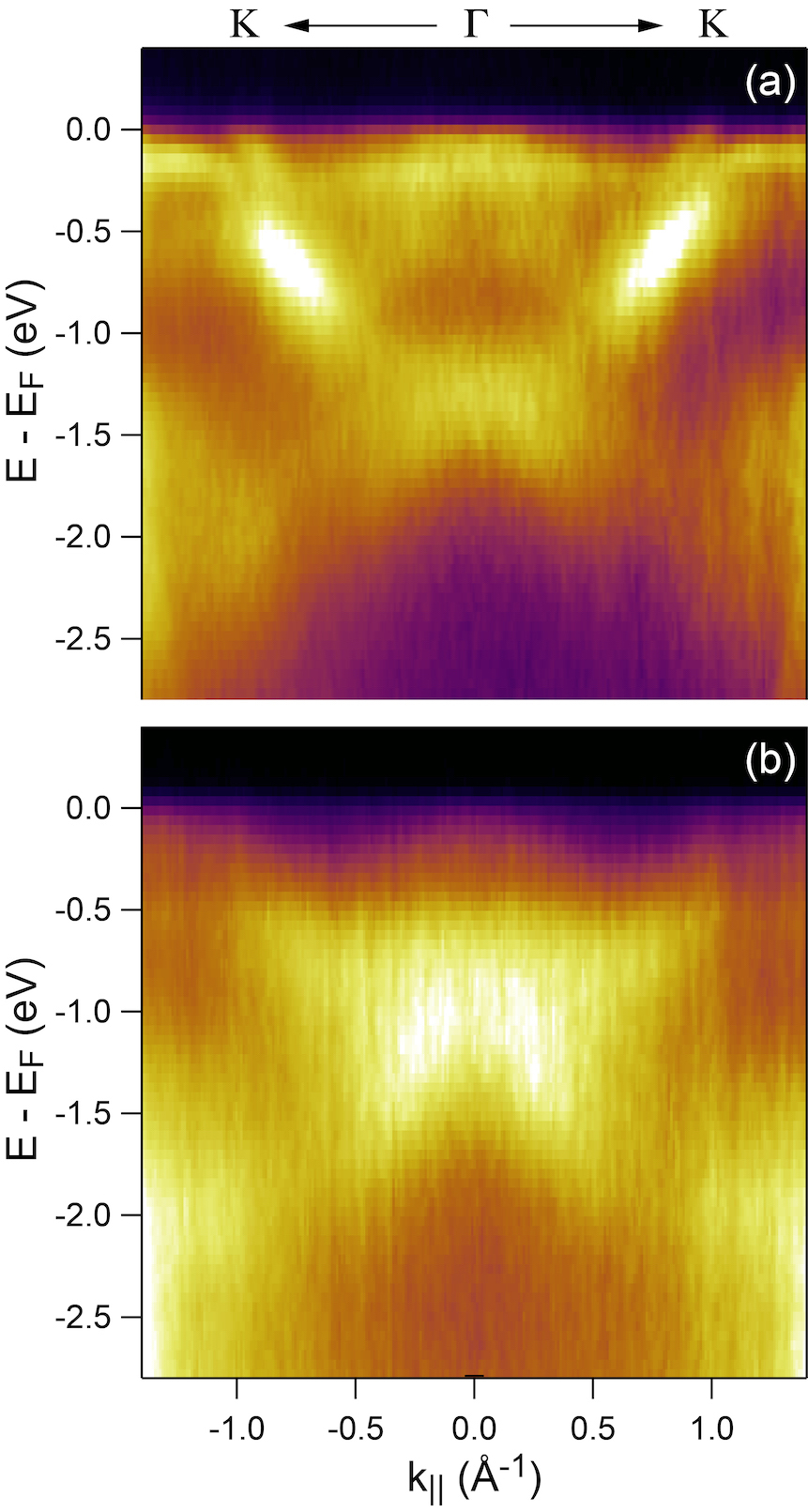}
	\caption{ARPES intensity maps zoomed around the $\Gamma$ point and $E_F$ for (a) gr/Ni(111) and (b) gr/Ni$_3$Mn/Ni(111).}
	\label{ARPES_GKzoom}
\end{figure}

Adsorption of graphene on Ni(111) strongly modifies the surface electronic structure of the Ni substrate and, as was shown, leads to the shift of the surface states of Ni(111) into the energy range above $E_F$~\cite{Achilli:2018if}. The 2D ARPES intensity map for gr/Ni(111) (Fig.~\ref{ARPES_GKzoom}(a)) shows undressed surface projections of the Ni energy bands: $d_\downarrow$ disperses from $E-E_F=-1.4$\,eV at $\Gamma$ to $E_F$ at $k_{||}\approx1$\,{\AA}$^{-1}$ and $sp_\downarrow$ crosses $E_F$ at $k_{||}\approx1$\,{\AA}$^{-1}$~\cite{Kreutz:1998aa,Mulazzi:2006fw}. After the formation of the thin layer of Ni$_3$Mn at the gr/Ni(111) interface, the electronic structure is modified as shown in Fig.~\ref{ARPES_GKzoom}(b): the intensity of the $d_\downarrow$ band is suppressed and the $sp_\downarrow$ is not present in the spectra. At the same time the intense bands at the $\Gamma$ point, which can be assigned to the Ni\,$3d$ states or to the mixture of the Ni+Mn\,$3d$ bands, are shifted from $E-E_F\approx-1.4$\,eV for gr/Ni(111) to $E-E_F\approx-1$\,eV for gr/Ni$_3$Mn/Ni(111), respectively~\cite{Kulkova:2002gn,Palumbo:2014ks}. All these changes, which clearly demonstrate the occurrence of a new phase (i.\,e. Ni$_3$Mn) as also shown by the structural data and discussed below, have implications on the electronic and transport properties of graphene in this system.   

\begin{figure}[t]
	\includegraphics[width=0.7\columnwidth]{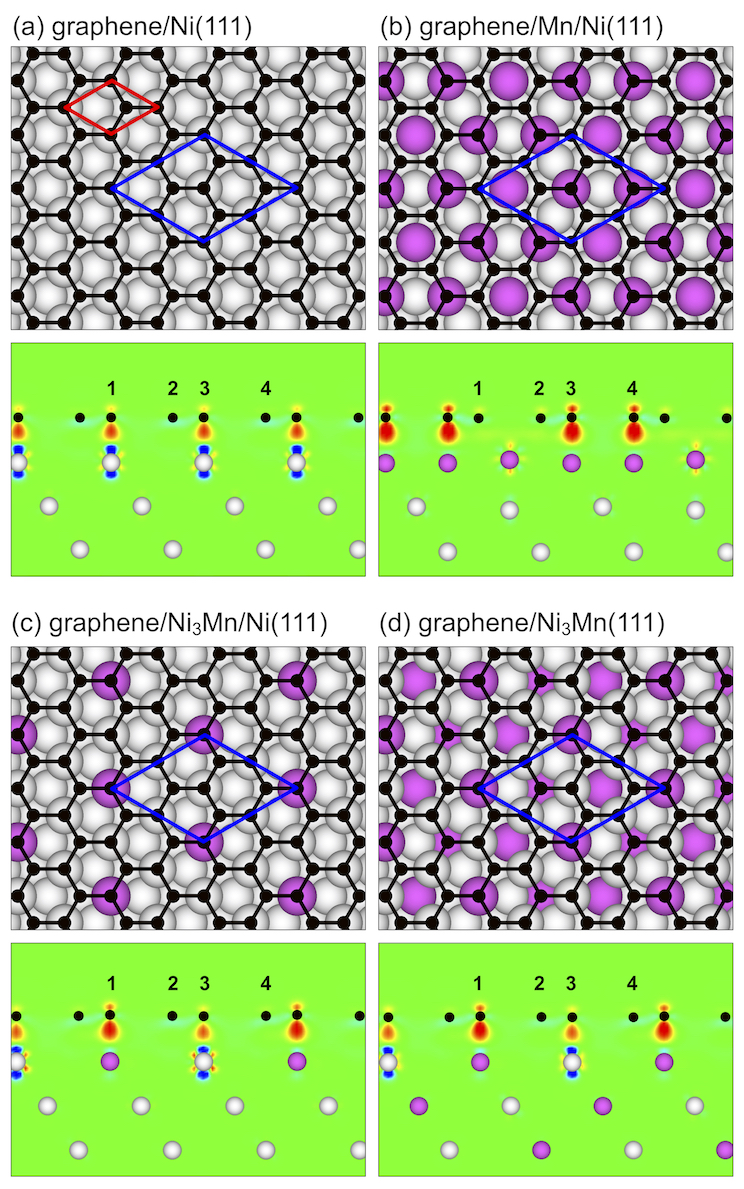}
	\caption{Top and side views of different Mn-based intercalation structures: (a) parent gr/Ni(111), (b) gr/$1$\,ML-Mn/Ni(111), (c) gr/$1$\,ML-Ni$_3$Mn/Ni(111), (d) gr/bulk-Ni$_3$Mn(111). Side views for all structures are taken along the main diagonal of the big rhombus marked in the top views and they are overlaid with electron charge difference maps $\Delta\rho(r)=\rho_{gr/s}-\rho_{gr}-\rho_{s}$ ($s$ - substrate). $\Delta\rho$ is colour coded as red ($+0.02e/\mathrm{\AA}^{3}$) -- green ($0$) -- blue ($-0.02e/\mathrm{\AA}^{3}$).}
	\label{grMnNi_structures_charges}
\end{figure}

\textit{DFT}. The experimental data were compared with the respective DFT results for the different Mn-intercalation gr/Ni(111) structures. The energetically most favourable structures after geometry optimization are presented in Fig.~\ref{grMnNi_structures_charges} (a-d). For the parent gr/Ni(111) system (Fig.~\ref{grMnNi_structures_charges} (a)) the so-called \textit{top-fcc} configuration of C-atoms above Ni(111) is the most energetically favourable with a distance between a graphene layer and top Ni layer of $2.09$\,\AA. In case of $1$\,ML-Mn layer placed between graphene and Ni(111) the Mn-Mn distance is $2.88$\,\AA, which is by $5$\% large compared to the $fcc$ $\gamma$-Mn stabilized on Pd(111)~\cite{Tian:1992aa} (Fig.~\ref{grMnNi_structures_charges} (b)). For the next two systems, a graphene layer is placed either on $1$\,ML-thick Ni$_3$Mn-layer on Ni(111) (Fig.~\ref{grMnNi_structures_charges} (c)) or on a bulk layer of Ni$_3$Mn (Fig.~\ref{grMnNi_structures_charges} (d)). The equilibrium distances between graphene and the top metallic layer in all structures are (b) $2.04$\,\AA, (c) $2.13$\,\AA, and (d) $2.12$\,\AA, respectively. 

Due to the small change of the distance between graphene and the underlying metallic support for all considered systems, before and after Mn intercalation in gr/Ni(111), an electron transfer from the underlying metal on graphene is observed, leading to $n$-doped graphene in all cases (see $\Delta \rho(r)$ maps in the side views of Fig.~\ref{grMnNi_structures_charges}). Due to the fact that the Dirac cone in all systems is fully destroyed due to the hybridization of the graphene $\pi$ and Ni(Mn) $3d$ states, the value of doping change after Mn intercalation in experimental and theoretical data can be compared with the shift of the position of the graphene $\pi$ band at the $\Gamma$ point measured for the systems before and after Mn intercalation in gr/Ni(111). The calculated shifts of this band are $0.48$\,eV for gr/$1$\,ML-Mn/Ni(111) (Fig.~\ref{grMnNi_structures_charges}(b)), $0.18$\,eV for gr/$1$\,ML-Ni$_3$Mn/Ni(111) (Fig.~\ref{grMnNi_structures_charges}(c)), and $0.24$\,eV for gr/Ni$_3$Mn(111) (Fig.~\ref{grMnNi_structures_charges}(d)), thus supporting our earlier conclusions about Ni$_3$Mn formation under the graphene layer. 

The intercalation of Mn in gr/Ni(111) leads to changes of the magnetic configuration in graphene layer compared to the parent system. The resulting magnetic moments of C atoms in the considered systems are presented in Tab.~\ref{magnetic_moments} (C$1$-C$4$ are referenced to marks in Fig.~\ref{grMnNi_structures_charges}). For gr/$1$\,ML-Mn/Ni(111), all magnetic moments in graphene are coupled antiferromagnetically with respect to the magnetization of the Ni(111) substrate, caused by the presence of the Mn layer at the interface. For the other two systems with Ni$_3$Mn underneath a graphene layer, the configurations of magnetic moments in graphene are close to each other - magnetic moments of the C atoms are aligned parallel to the magnetization of the Ni(111) bulk or Ni$_3$Mn, except for C1. However, it is interesting to note that the total absolute magnetization of graphene (ferromagnetically or antiferromagnetically coupled to Ni(111)) is increased for all Mn-intercalation systems compared to gr/Ni(111) allowing the effective spin manipulation in graphene which is in contact with such Ni-Mn interfaces.

\begin{figure}[t]
	\includegraphics[width=\columnwidth]{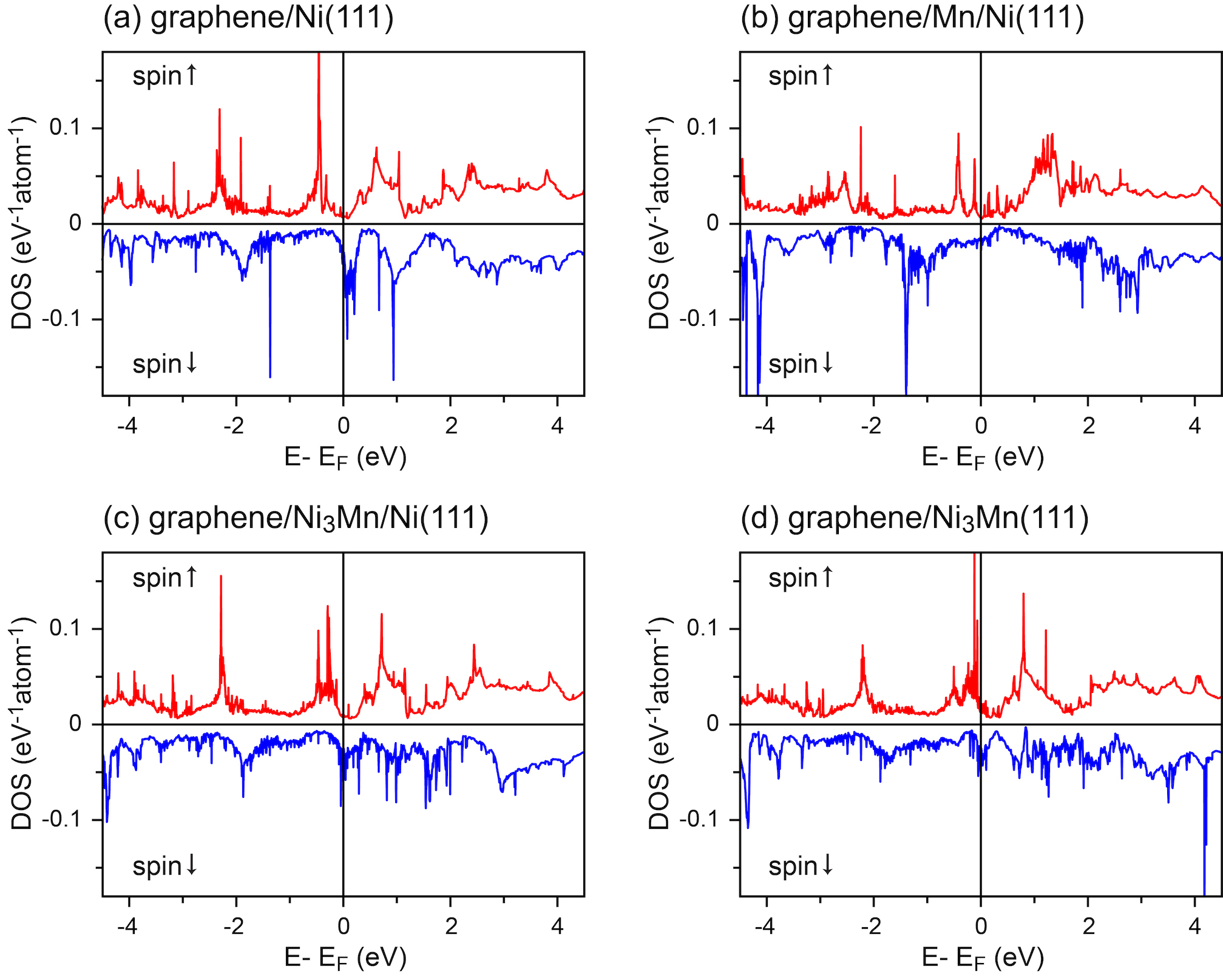}
	\caption{Spin-resolved C-$p_z$-projected density of states for (a) gr/Ni(111), (b) gr/$1$\,ML-Mn/Ni(111), (c) gr/$1$\,ML-Ni$_3$Mn/Ni(111), and (d) gr/bulk-Ni$_3$Mn(111).}
	\label{grMnNi_DOS}
\end{figure}

\begin{figure}[t]
	\includegraphics[width=0.9\columnwidth]{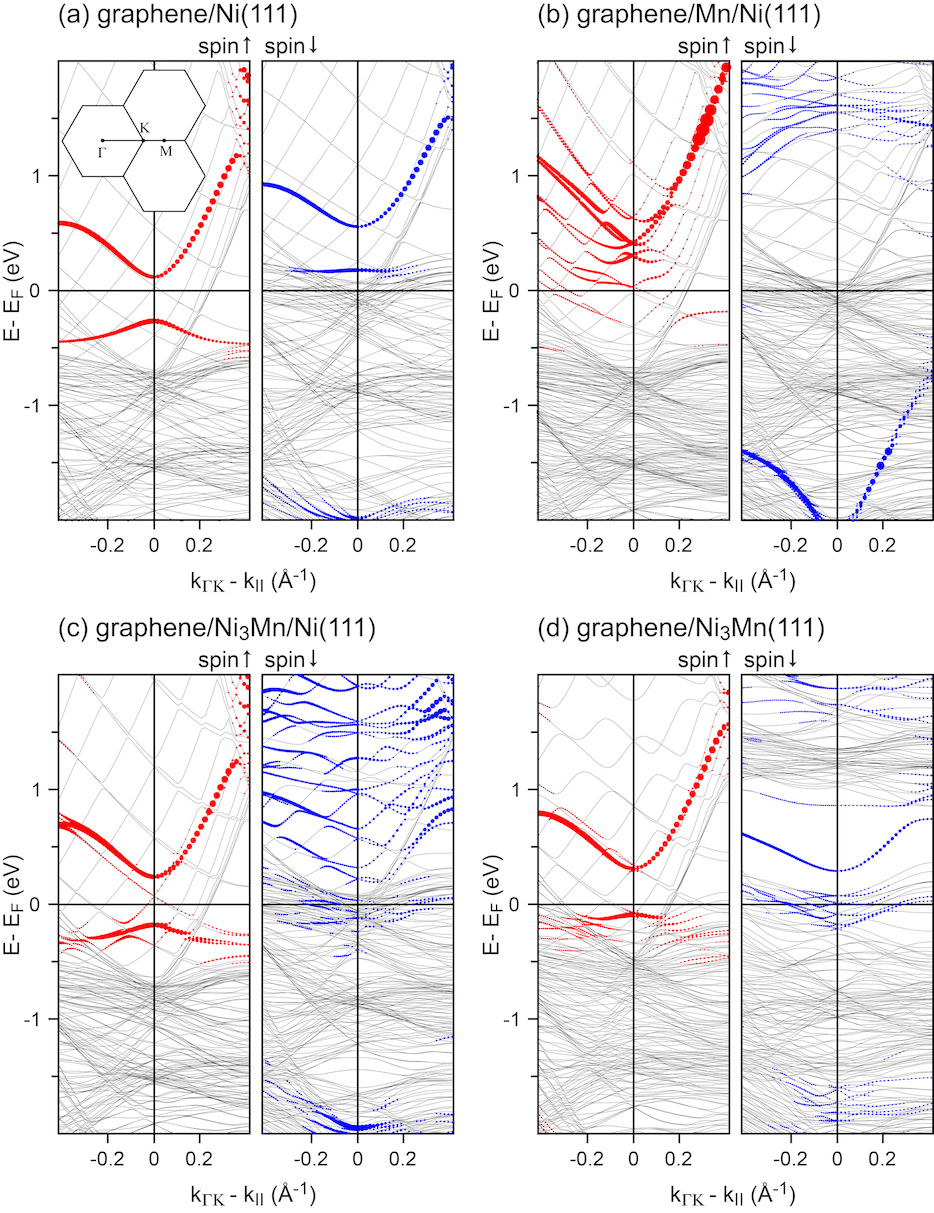}
	\caption{Spin-resolved band structures for (a) gr/Ni(111), (b) gr/$1$\,ML-Mn/Ni(111), (c) gr/$1$\,ML-Ni$_3$Mn/Ni(111), and (d) gr/bulk-Ni$_3$Mn(111). Band structures are presented around $\Gamma$ point for the graphene derived $(1\times1)$ Brillouin zone. The weight of the graphene-derived $p_z$ character is highlighted by the size of filled circles superimposed with the plot of the band structure.}
	\label{grMnNi_bands}
\end{figure}

Figures~\ref{grMnNi_DOS} and ~\ref{grMnNi_bands} show the calculated spin-resolved partial density of states for graphene $\pi$ states and the spin-resolved band structure, respectively, for the systems: (a) gr/Ni(111), (b) gr/$1$\,ML-Mn/Ni(111), (c) gr/$1$\,ML-Ni$_3$Mn/Ni(111), and (d) gr/bulk-Ni$_3$Mn(111). For a more accurate comparison, the unit cell of gr/Ni(111) was modelled in the same $(2\times2)$ periodicity as for Mn-based interfaces. The calculated band structure for gr/Ni(111) is similar to the one presented in earlier works~\cite{Weser:2011,Voloshina:2011NJP} except for the folded bands originating from the doubled periodicity of the system.

Comparing our NEXAFS and ARPES experimental data with the presented band structures we can find a close similarity between experimental and theoretical band structures for gr/Ni(111) and gr/Ni$_3$Mn/Ni(111). [In the present situation it is very difficult (if not impossible) to estimate from the experimental data the thickness of the Ni$_3$Mn layer. However, the presented results (Fig.~\ref{grMnNi_bands}(c,d)) show that the electronic structures of both gr/Ni$_3$Mn interfaces are very similar.] Intercalation of Mn and the formation of Ni$_3$Mn underneath of a graphene layer lead to a small upward shift of the graphene-derived spin-up bands by $\approx100$\,meV leaving the value of the gap between upper and lower branches of $\approx400$\,meV almost unchanged. At the same time, the structure of the interface states for both spins between graphene and the metallic support becomes smeared in energy due to the hybridization between the respective valence band states of graphene and the Ni$_3$Mn alloy. All these facts are in good agreement with experimental data where a smaller $n$-doping of graphene and suppression of the intensity of the interface states around the $\mathrm{K}$ point was observed. 

Close analysis of the calculated band structure for gr/$1$\,ML-Ni$_3$Mn/Ni(111), which we conclude to be formed in the present experiment, shows that a Dirac-like energy bands crossing is formed in the energy gap for spin-up states above the Fermi level at a binding energy $E-E_F=+0.072$\,eV. The orbital analysis of this state demonstrates the contribution of the graphene $\pi$ character with the dominant contribution of the Ni\,$3d$ states. Further analysis shows that the considered bands are not the result of band folding; obviously a deeper analysis of the nature of these states is necessary.

Finally, it is interesting to consider the calculated electronic structure for the sharp interface gr/$1$\,ML-Mn/Ni(111) (Fig.~\ref{grMnNi_bands}(b)). For spin-down states, a very big $n$-doping of graphene is observed and the Dirac point in this case is located at $E-E_F=-2.3$\,eV. A similar situation is also found for the spin-up states although the structure of the Dirac point is less pronounced. This situation is not obvious for the case of graphene adsorption on a metal with a partially filled $d$ shell (Mn has an electronic configuration $3d^54s^2$). Here one can speculate that the half-filled $3d$ orbital has an orbital moment of $L=0$ which leads to the spherical symmetry of atomic orbitals in Mn. Therefore Mn might behave like Ca donating electrons to graphene and producing a relatively large $n$-doping of the graphene layer at a very short distance of $2.04$\,\AA\ between graphene and Mn. This property of the Mn-intercalated layer can be used in further experiments on intercalation of Mn layers underneath graphene where the formation of Mn-alloys can be avoided.

\section{Conclusions}

The intercalation of Mn atoms in the gr/Ni(111) interface leads to the formation of a graphene-protected thin Ni$_3$Mn layer as deduced from our experimental data. At the gr/Ni$_3$Mn interface, the graphene layer exhibits a slightly reduced $n$-doping compared to the parent gr/Ni(111) system. All experimental results are analyzed in the framework of state-of-the-art DFT calculations, demonstrating very good agreement between all data. The observed effects are explained by the hybridization between the valence band states of graphene and the underlying strong FM Ni$_3$Mn. Additionally, the electronic structures of different gr/Mn-Ni interfaces were analysed with DFT, and we conclude that a hypothetical sharp gr-Mn interface prepared on a substrate without alloying effects would lead to strongly $n$-doped graphene, which can be used in further studies of different physical phenomena in graphene-based low-dimensional systems. 

\section*{Experimental}

All experiments were performed in several stations under identical experimental conditions allowing reproducibility of all results. We prepared gr/Ni(111) under UHV conditions following the procedure described in the literature: a pre-cleaned Ni(111) crystal was exposed to C$_2$H$_4$ at $600^\circ$\,C and $p=1\times10^{-6}$\,mbar for $15$\,min~\cite{Dedkov:2017jn}. Manganese was evaporated from a carefully degassed $e$-beam source at a pressure below $p=5\times10^{-10}$\,mbar and its thickness (slightly more than $1$\,ML) was measured with a quartz microbalance and then further estimated on the basis of spectroscopic experiments. Annealing of the Mn/gr/Ni(111) system at $400^\circ$\,C leads to the formation of the ordered gr/Ni$_3$Mn interface. The ordering and cleanness of the system at every preparation step was verified with low-energy electron diffraction (LEED), scanning tunnelling microscopy (STM) and x-ray photoelectron spectroscopy (XPS).

Initial preparation/spectroscopic experiments were performed in the laboratory station equipped with a SPECS PHOIBOS\,150 hemispherical analyzer and combined non-monochromotized Al\,K$\alpha$/Mg\,K$\alpha$ x-ray source. Further angle-resolved photoelectron spectroscopy (ARPES) experiments were performed at the UE\,56/2-PGM beamline of the BESSY\,II storage ring (HZB Berlin). ARPES data were collected in the snapshot mode with a SPECS PHOIBOS\,100 hemispherical analyzer equipped with the 2D-CCD detector while the sample was placed on a 5-axis fully-motorized manipulator. The sample was pre-aligned (via polar and azimuth angles rotations) in such a way that the tilt scan was performed along the $\Gamma-\mathrm{K}$ direction of the graphene-derived Brillouin zone (BZ) with the photoemission intensity on the channelplate images acquired along the direction perpendicular to $\Gamma-\mathrm{K}$. The final 3D data set of the photoemission intensity as a function of kinetic energy and two emission angles, $I(E_{kin};angle1;angle2)$, were then carefully analyzed. NEXAFS data were collected at the D1011 beamline of MAX-lab (Lund) in the partial and total electron yield modes (PEY and TEY).

STM experiments were performed in the lab-station equipped with a SPECS STM Aarhus 150. STM measurements were performed in constant current mode at room temperature using electrochemically etched polycrystalline tungsten tips cleaned in UHV via Ar$^+$ sputtering. The sign of the bias voltage corresponds to the voltage applied to the sample. Tunnelling current and voltage are labelled $I_T$ and $U_T$, respectively. In this station the additional XPS equipment (SPECS PHOIBOS\,150 and SPECS monochromotized Al\,K$\alpha$/Ag\,L$\alpha$ x-ray source) was used for the control experiments during sample preparation. 

DFT calculations based on plane-wave basis sets of $500$\,eV cutoff energy were performed with the Vienna \textit{ab initio} simulation package (VASP)~\cite{Kresse:1994,Kresse:1996a}. The Perdew-Burke-Ernzerhof (PBE) exchange-correlation functional~\cite{Perdew:1996} was employed. The electron-ion interaction was described within the projector augmented wave (PAW) method~\cite{Blochl:1994} with C ($2s$, $2p$), Ni ($3d$, $4s$), Mn ($3d$, $4s$) states treated as valence states. The BZ integration was performed on $\Gamma$-centred symmetry reduced Monkhorst-Pack meshes using a Methfessel-Paxton smearing method of first order with $\sigma = 0.15$\,eV, except for the calculation of total energies. For these calculations, the tetrahedron method with Bl\"ochl corrections~\cite{Blochl:1994vg} was employed. The $k$-mesh for sampling of the supercell Brillouin zone was chosen to be as dense as $24\times24$, when folded up to the simple graphene unit cell. Dispersion interactions were considered adding a $1/r^6$ atom-atom term as parameterised by Grimme (``D2'' parameterisation)~\cite{Grimme:2006}.

The studied systems are modelled using supercells which have a ($2\times2$) lateral periodicity with respect to graphene and contain $13$ layers of metal atoms ($4$ atoms per layer), a graphene sheet ($8$ atoms per layer) adsorbed on both sides of the slab and a vacuum gap of approximately $23$\,\AA. In the case of the graphene-Mn-Ni(111) system, an additional Mn-layer is added from both sides of the slab between graphene and Ni(111). During the structural optimisation the positions of the carbon atoms ($x$,$y$,$z$-coordinates) as well as those of the top three layers of metal atoms ($z$-coordinates) are relaxed until forces became smaller than $0.02\,\mathrm{eV\AA}^{-1}$.

The band structures calculated for the studied systems were unfolded (if necessary) to the graphene ($1\times1$) primitive unit cells according to the procedure described in Ref.~\citenum{Medeiros:2014ka} and \citenum{Medeiros:2015ks} with the code BandUP. 


\section*{Acknowledgement}

The computing facilities of the Freie Universit\"at Berlin (ZEDAT) and the North-German Supercomputing Alliance (HLRN) are acknowledged for computer time. Yi Cui is grateful to the support from Natural Science Foundation of Jiangsu Province (BK20170426). We acknowledge Alexei Preobrajenski for his help during beamtime at MAX-lab.

\providecommand{\latin}[1]{#1}
\makeatletter
\providecommand{\doi}
  {\begingroup\let\do\@makeother\dospecials
  \catcode`\{=1 \catcode`\}=2 \doi@aux}
\providecommand{\doi@aux}[1]{\endgroup\texttt{#1}}
\makeatother
\providecommand*\mcitethebibliography{\thebibliography}
\csname @ifundefined\endcsname{endmcitethebibliography}
  {\let\endmcitethebibliography\endthebibliography}{}


\clearpage
\begin{table}
\caption{The interface carbon spin magnetic moments (in $\mu_B$) as computed for the atoms marked in Figure~\ref{grMnNi_structures_charges}.}
\label{magnetic_moments}
\begin{tabular}{p{4cm} p{2cm} p{2cm} p{2cm} p{2cm} }
\hline
System & C1 & C2 & C3 & C4\\
\hline
(a) gr/Ni(111)         &$-0.019$ &$0.034$ &$-0.019$ &$0.034$ \\[0.5cm]
(b) gr/Mn/Ni(111)      &$-0.018$ &$-0.018$ &$-0.029$ &$-0.029$\\[0.5cm]
(c) gr/Ni$_3$Mn/Ni(111)&$-0.047$ & $0.021$ & $0.050$ & $0.021$\\[0.5cm]
(d) gr/Ni$_3$Mn(111)   &$-0.051$ &$0.023$ & $0.037$ & $0.023$\\
\hline
\end{tabular}
\end{table}

\clearpage

\clearpage

\clearpage

\clearpage

\clearpage

\clearpage

\clearpage

\clearpage
\begin{figure}[t]
	\includegraphics[width=0.9\columnwidth]{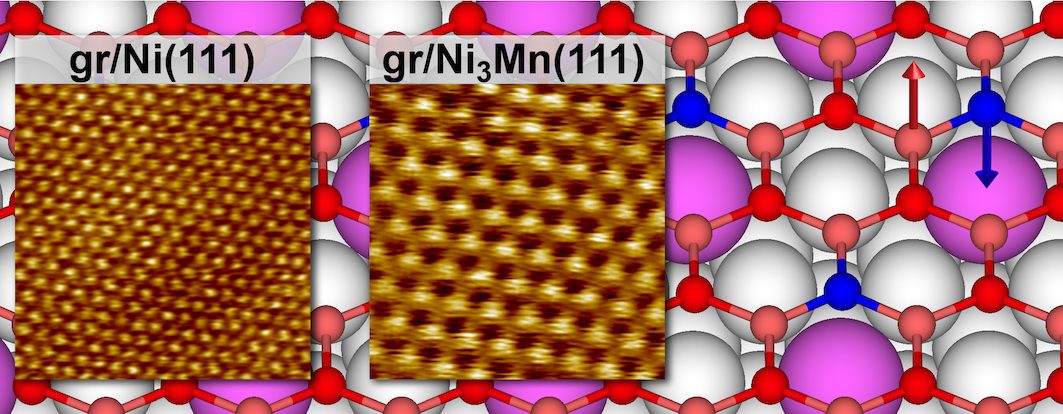}
\end{figure}


\begin{mcitethebibliography}{55}
\providecommand*\natexlab[1]{#1}
\providecommand*\mciteSetBstSublistMode[1]{}
\providecommand*\mciteSetBstMaxWidthForm[2]{}
\providecommand*\mciteBstWouldAddEndPuncttrue
  {\def\EndOfBibitem{\unskip.}}
\providecommand*\mciteBstWouldAddEndPunctfalse
  {\let\EndOfBibitem\relax}
\providecommand*\mciteSetBstMidEndSepPunct[3]{}
\providecommand*\mciteSetBstSublistLabelBeginEnd[3]{}
\providecommand*\EndOfBibitem{}
\mciteSetBstSublistMode{f}
\mciteSetBstMaxWidthForm{subitem}{(\alph{mcitesubitemcount})}
\mciteSetBstSublistLabelBeginEnd
  {\mcitemaxwidthsubitemform\space}
  {\relax}
  {\relax}

\bibitem[Batzill(2012)]{Batzill:2012}
Batzill,~M. The Surface Science of Graphene: Metal Interfaces, CVD Synthesis, Nanoribbons, Chemical Modifications, and Defects. \emph{Surf. Sci. Rep.} \textbf{2012}, \emph{67}, 83--115\relax
\mciteBstWouldAddEndPuncttrue
\mciteSetBstMidEndSepPunct{\mcitedefaultmidpunct}
{\mcitedefaultendpunct}{\mcitedefaultseppunct}\relax
\EndOfBibitem
\bibitem[Dedkov and Voloshina(2015)Dedkov, and Voloshina]{Dedkov:2015kp}
Dedkov,~Y.; Voloshina,~E. Graphene Growth and Properties on Metal Substrates. \emph{J. Phys.: Condens. Matter} \textbf{2015},
  \emph{27}, 303002\relax
\mciteBstWouldAddEndPuncttrue
\mciteSetBstMidEndSepPunct{\mcitedefaultmidpunct}
{\mcitedefaultendpunct}{\mcitedefaultseppunct}\relax
\EndOfBibitem
\bibitem[Bae \latin{et~al.}(2010)Bae, Kim, Lee, Xu, Park, Zheng, Balakrishnan,
  Lei, Kim, Song, Kim, Kim, Ozyilmaz, Ahn, Hong, and Iijima]{Bae:2010}
Bae,~S. \latin{et~al.}  Roll-to-Roll Production of 30-Inch Graphene Films
for Transparent Electrodes. \emph{Nat. Nanotech.} \textbf{2010}, \emph{5},
  574--578\relax
\mciteBstWouldAddEndPuncttrue
\mciteSetBstMidEndSepPunct{\mcitedefaultmidpunct}
{\mcitedefaultendpunct}{\mcitedefaultseppunct}\relax
\EndOfBibitem
\bibitem[Liu \latin{et~al.}(2011)Liu, Fu, Dai, Yan, Liu, Zhao, Zhang, and
  Liu]{Liu:2011bl}
Liu,~N.; Fu,~L.; Dai,~B.; Yan,~K.; Liu,~X.; Zhao,~R.; Zhang,~Y.; Liu,~Z. Universal Segregation Growth Approach to Wafer-Size Graphene from Non-Noble Metals.
  \emph{Nano Lett.} \textbf{2011}, \emph{11}, 297--303\relax
\mciteBstWouldAddEndPuncttrue
\mciteSetBstMidEndSepPunct{\mcitedefaultmidpunct}
{\mcitedefaultendpunct}{\mcitedefaultseppunct}\relax
\EndOfBibitem
\bibitem[Lee \latin{et~al.}(2017)Lee, Liu, Chai, Mohamed, Aziz, Khe, Hidayah,
  and Hashim]{Lee:2017cm}
Lee,~H.~C.; Liu,~W.-W.; Chai,~S.-P.; Mohamed,~A.~R.; Aziz,~A.; Khe,~C.-S.;
  Hidayah,~N. M.~S.; Hashim,~U. Review of the Synthesis, Transfer, Characterization and Growth Mechanisms of Single and Multilayer Graphene. \emph{RSC Adv.} \textbf{2017}, \emph{7},
  15644--15693\relax
\mciteBstWouldAddEndPuncttrue
\mciteSetBstMidEndSepPunct{\mcitedefaultmidpunct}
{\mcitedefaultendpunct}{\mcitedefaultseppunct}\relax
\EndOfBibitem
\bibitem[Kim \latin{et~al.}(2009)Kim, Zhao, Jang, Lee, Kim, Kim, Ahn, Kim,
  Choi, and Hong]{Kim:2009a}
Kim,~K.~S.; Zhao,~Y.; Jang,~H.; Lee,~S.~Y.; Kim,~J.~M.; Kim,~K.~S.; Ahn,~J.-H.;
  Kim,~P.; Choi,~J.-Y.; Hong,~B.~H. Large-Scale Pattern Growth of Graphene Films for Stretchable Transparent Electrodes. \emph{Nature} \textbf{2009}, \emph{457},
  706--710\relax
\mciteBstWouldAddEndPuncttrue
\mciteSetBstMidEndSepPunct{\mcitedefaultmidpunct}
{\mcitedefaultendpunct}{\mcitedefaultseppunct}\relax
\EndOfBibitem
\bibitem[Ryu \latin{et~al.}(2014)Ryu, Kim, Won, Kim, Park, Lee, Cho, Cho, Kim,
  Ryu, Shin, Lee, Hong, and Cho]{Ryu:2014fo}
Ryu,~J.; Kim,~Y.; Won,~D.; Kim,~N.; Park,~J.~S.; Lee,~E.-K.; Cho,~D.;
  Cho,~S.-P.; Kim,~S.~J.; Ryu,~G.~H.; Shin,~H.-A.-S.; Lee,~Z.; Hong,~B.~H.;
  Cho,~S. Fast Synthesis of High-Performance
Graphene Films by Hydrogen-Free Rapid
Thermal Chemical Vapor Deposition. \emph{ACS Nano} \textbf{2014}, \emph{8}, 950--956\relax
\mciteBstWouldAddEndPuncttrue
\mciteSetBstMidEndSepPunct{\mcitedefaultmidpunct}
{\mcitedefaultendpunct}{\mcitedefaultseppunct}\relax
\EndOfBibitem
\bibitem[Dedkov \latin{et~al.}(2008)Dedkov, Fonin, and Laubschat]{Dedkov:2008d}
Dedkov,~Y.~S.; Fonin,~M.; Laubschat,~C. A Possible Source of Spin-Polarized Electrons: The Inert Graphene/Ni(111)
system. \emph{Appl. Phys. Lett.} \textbf{2008},
  \emph{92}, 052506\relax
\mciteBstWouldAddEndPuncttrue
\mciteSetBstMidEndSepPunct{\mcitedefaultmidpunct}
{\mcitedefaultendpunct}{\mcitedefaultseppunct}\relax
\EndOfBibitem
\bibitem[Dedkov \latin{et~al.}(2008)Dedkov, Fonin, Ruediger, and
  Laubschat]{Dedkov:2008e}
Dedkov,~Y.~S.; Fonin,~M.; Ruediger,~U.; Laubschat,~C. Graphene-Protected Iron Layer on Ni(111). \emph{Appl. Phys. Lett.}
  \textbf{2008}, \emph{93}, 022509\relax
\mciteBstWouldAddEndPuncttrue
\mciteSetBstMidEndSepPunct{\mcitedefaultmidpunct}
{\mcitedefaultendpunct}{\mcitedefaultseppunct}\relax
\EndOfBibitem
\bibitem[Sutter \latin{et~al.}(2010)Sutter, Albrecht, Camino, and
  Sutter]{Sutter:2010bx}
Sutter,~E.; Albrecht,~P.; Camino,~F.~E.; Sutter,~P. Monolayer Graphene as Ultimate Chemical Passivation Layer for Arbitrarily Shaped Metal Surfaces. \emph{Carbon}
  \textbf{2010}, \emph{48}, 4414--4420\relax
\mciteBstWouldAddEndPuncttrue
\mciteSetBstMidEndSepPunct{\mcitedefaultmidpunct}
{\mcitedefaultendpunct}{\mcitedefaultseppunct}\relax
\EndOfBibitem
\bibitem[Weatherup \latin{et~al.}(2015)Weatherup, D'Arsie, Cabrero-Vilatela,
  Caneva, Blume, Robertson, Schloegl, and Hofmann]{Weatherup:2015cx}
Weatherup,~R.~S.; D'Arsie,~L.; Cabrero-Vilatela,~A.; Caneva,~S.; Blume,~R.;
  Robertson,~J.; Schloegl,~R.; Hofmann,~S. Long-Term Passivation of Strongly Interacting Metals with Single-Layer Graphene. \emph{J. Am. Chem. Soc.}
  \textbf{2015}, \emph{137}, 14358--14366\relax
\mciteBstWouldAddEndPuncttrue
\mciteSetBstMidEndSepPunct{\mcitedefaultmidpunct}
{\mcitedefaultendpunct}{\mcitedefaultseppunct}\relax
\EndOfBibitem
\bibitem[Campbell \latin{et~al.}(2018)Campbell, Kiraly, Jacobberger, Mannix,
  Arnold, Hersam, Guisinger, and Bedzyk]{Campbell:2018ii}
Campbell,~G.~P.; Kiraly,~B.; Jacobberger,~R.~M.; Mannix,~A.~J.; Arnold,~M.~S.;
  Hersam,~M.~C.; Guisinger,~N.~P.; Bedzyk,~M.~J. Epitaxial Graphene-Encapsulated Surface Reconstruction of Ge(110). \emph{Phys. Rev. Mater.}
  \textbf{2018}, \emph{2}, 044004\relax
\mciteBstWouldAddEndPuncttrue
\mciteSetBstMidEndSepPunct{\mcitedefaultmidpunct}
{\mcitedefaultendpunct}{\mcitedefaultseppunct}\relax
\EndOfBibitem
\bibitem[Tesch \latin{et~al.}(2018)Tesch, Paschke, Fonin, Wietstruk,
  B{\"o}ttcher, Koch, Bostwick, Jozwiak, Rotenberg, Makarova, Paulus,
  Voloshina, and Dedkov]{Tesch:2018hm}
Tesch,~J.; Paschke,~F.; Fonin,~M.; Wietstruk,~M.; B{\"o}ttcher,~S.;
  Koch,~R.~J.; Bostwick,~A.; Jozwiak,~C.; Rotenberg,~E.; Makarova,~A.;
  Paulus,~B.; Voloshina,~E.; Dedkov,~Y. The Graphene/n-Ge(110) Interface: Structure, Doping, and Electronic Properties. \emph{Nanoscale} \textbf{2018},
  \emph{10}, 6088--6098\relax
\mciteBstWouldAddEndPuncttrue
\mciteSetBstMidEndSepPunct{\mcitedefaultmidpunct}
{\mcitedefaultendpunct}{\mcitedefaultseppunct}\relax
\EndOfBibitem
\bibitem[Zhou \latin{et~al.}(2018)Zhou, Niu, and Niu]{Zhou:2018ji}
Zhou,~D.; Niu,~Z.; Niu,~T. Surface Reconstruction of Germanium:
Hydrogen Intercalation and Graphene Protection. \emph{J. Phys. Chem. C}
  \textbf{2018}, \emph{122}, 21874--21882\relax
\mciteBstWouldAddEndPuncttrue
\mciteSetBstMidEndSepPunct{\mcitedefaultmidpunct}
{\mcitedefaultendpunct}{\mcitedefaultseppunct}\relax
\EndOfBibitem
\bibitem[Voloshina and Dedkov(2012)Voloshina, and Dedkov]{Voloshina:2012c}
Voloshina,~E.; Dedkov,~Y. Graphene on Metallic Surfaces: Problems and Perspectives. \emph{Phys. Chem. Chem. Phys.} \textbf{2012},
  \emph{14}, 13502--13514\relax
\mciteBstWouldAddEndPuncttrue
\mciteSetBstMidEndSepPunct{\mcitedefaultmidpunct}
{\mcitedefaultendpunct}{\mcitedefaultseppunct}\relax
\EndOfBibitem
\bibitem[Voloshina and Dedkov(2014)Voloshina, and Dedkov]{Voloshina:2014jl}
Voloshina,~E.~N.; Dedkov,~Y.~S. General Approach to Understanding the Electronic Structure of Graphene on Metals. \emph{Mater. Res. Express}
  \textbf{2014}, \emph{1}, 035603\relax
\mciteBstWouldAddEndPuncttrue
\mciteSetBstMidEndSepPunct{\mcitedefaultmidpunct}
{\mcitedefaultendpunct}{\mcitedefaultseppunct}\relax
\EndOfBibitem
\bibitem[Karpan \latin{et~al.}(2007)Karpan, Giovannetti, Khomyakov, Talanana,
  Starikov, Zwierzycki, Brink, Brocks, and Kelly]{Karpan:2007}
Karpan,~V.~M.; Giovannetti,~G.; Khomyakov,~P.~A.; Talanana,~M.;
  Starikov,~A.~A.; Zwierzycki,~M.; Brink,~J. v.~d.; Brocks,~G.; Kelly,~P.~J. Graphite and Graphene as Perfect Spin Filters.
  \emph{Phys. Rev. Lett.} \textbf{2007}, \emph{99}, 176602\relax
\mciteBstWouldAddEndPuncttrue
\mciteSetBstMidEndSepPunct{\mcitedefaultmidpunct}
{\mcitedefaultendpunct}{\mcitedefaultseppunct}\relax
\EndOfBibitem
\bibitem[Karpan \latin{et~al.}(2008)Karpan, Khomyakov, Starikov, Giovannetti,
  Zwierzycki, Talanana, Brocks, Brink, and Kelly]{Karpan:2008}
Karpan,~V.~M.; Khomyakov,~P.~A.; Starikov,~A.~A.; Giovannetti,~G.;
  Zwierzycki,~M.; Talanana,~M.; Brocks,~G.; Brink,~J. v.~d.; Kelly,~P.~J. Theoretical Prediction of Perfect Spin Filtering at Interfaces Between Close-Packed Surfaces
of Ni or Co and Graphite or Graphene.
  \emph{Phys. Rev. B} \textbf{2008}, \emph{78}, 195419\relax
\mciteBstWouldAddEndPuncttrue
\mciteSetBstMidEndSepPunct{\mcitedefaultmidpunct}
{\mcitedefaultendpunct}{\mcitedefaultseppunct}\relax
\EndOfBibitem
\bibitem[Karpan \latin{et~al.}(2011)Karpan, Khomyakov, Giovannetti, Starikov,
  and Kelly]{Karpan:2011kv}
Karpan,~V.; Khomyakov,~P.; Giovannetti,~G.; Starikov,~A.; Kelly,~P. Ni(111)|graphene|h-BN Junctions as Ideal Spin Injectors. \emph{Phys.
  Rev. B} \textbf{2011}, \emph{84}\relax
\mciteBstWouldAddEndPuncttrue
\mciteSetBstMidEndSepPunct{\mcitedefaultmidpunct}
{\mcitedefaultendpunct}{\mcitedefaultseppunct}\relax
\EndOfBibitem
\bibitem[Weser \latin{et~al.}(2010)Weser, Rehder, Horn, Sicot, Fonin,
  Preobrajenski, Voloshina, Goering, and Dedkov]{Weser:2010}
Weser,~M.; Rehder,~Y.; Horn,~K.; Sicot,~M.; Fonin,~M.; Preobrajenski,~A.~B.;
  Voloshina,~E.~N.; Goering,~E.; Dedkov,~Y.~S. Induced Magnetism of Carbon Atoms at the Graphene/Ni(111) Interface. \emph{Appl. Phys. Lett.}
  \textbf{2010}, \emph{96}, 012504\relax
\mciteBstWouldAddEndPuncttrue
\mciteSetBstMidEndSepPunct{\mcitedefaultmidpunct}
{\mcitedefaultendpunct}{\mcitedefaultseppunct}\relax
\EndOfBibitem
\bibitem[Dedkov and Fonin(2010)Dedkov, and Fonin]{Dedkov:2010jh}
Dedkov,~Y.~S.; Fonin,~M. Electronic and Magnetic Properties of the Graphene-Ferromagnet Interface. \emph{New J. Phys.} \textbf{2010}, \emph{12},
  125004\relax
\mciteBstWouldAddEndPuncttrue
\mciteSetBstMidEndSepPunct{\mcitedefaultmidpunct}
{\mcitedefaultendpunct}{\mcitedefaultseppunct}\relax
\EndOfBibitem
\bibitem[Weser \latin{et~al.}(2011)Weser, Voloshina, Horn, and
  Dedkov]{Weser:2011}
Weser,~M.; Voloshina,~E.~N.; Horn,~K.; Dedkov,~Y.~S. Electronic Structure and Magnetic Properties of the Graphene/Fe/Ni(111) Intercalation-Like System. \emph{Phys. Chem. Chem.
  Phys.} \textbf{2011}, \emph{13}, 7534--7539\relax
\mciteBstWouldAddEndPuncttrue
\mciteSetBstMidEndSepPunct{\mcitedefaultmidpunct}
{\mcitedefaultendpunct}{\mcitedefaultseppunct}\relax
\EndOfBibitem
\bibitem[Marchenko \latin{et~al.}(2015)Marchenko, Varykhalov, Sanchez-Barriga,
  Rader, Carbone, and Bihlmayer]{Marchenko:2015ka}
Marchenko,~D.; Varykhalov,~A.; Sanchez-Barriga,~J.; Rader,~O.; Carbone,~C.;
  Bihlmayer,~G. Highly Spin-Polarized Dirac Fermions at the Graphene/Co Interface. \emph{Phys. Rev. B} \textbf{2015}, \emph{91}, 235431\relax
\mciteBstWouldAddEndPuncttrue
\mciteSetBstMidEndSepPunct{\mcitedefaultmidpunct}
{\mcitedefaultendpunct}{\mcitedefaultseppunct}\relax
\EndOfBibitem
\bibitem[Usachov \latin{et~al.}(2015)Usachov, Fedorov, Otrokov, Chikina,
  Vilkov, Petukhov, Rybkin, Koroteev, Chulkov, Adamchuk, Gr{\"u}neis,
  Laubschat, and Vyalikh]{Usachov:2015kr}
Usachov,~D.; Fedorov,~A.; Otrokov,~M.~M.; Chikina,~A.; Vilkov,~O.;
  Petukhov,~A.; Rybkin,~A.~G.; Koroteev,~Y.~M.; Chulkov,~E.~V.;
  Adamchuk,~V.~K.; Gr{\"u}neis,~A.; Laubschat,~C.; Vyalikh,~D.~V. 
Observation of Single-Spin Dirac Fermions at the Graphene/ Ferromagnet Interface. \emph{Nano
  Lett.} \textbf{2015}, \emph{15}, 2396--2401\relax
\mciteBstWouldAddEndPuncttrue
\mciteSetBstMidEndSepPunct{\mcitedefaultmidpunct}
{\mcitedefaultendpunct}{\mcitedefaultseppunct}\relax
\EndOfBibitem
\bibitem[Dedkov \latin{et~al.}(2001)Dedkov, Shikin, Adamchuk, Molodtsov,
  Laubschat, Bauer, and Kaindl]{Dedkov:2001}
Dedkov,~Y.~S.; Shikin,~A.~M.; Adamchuk,~V.~K.; Molodtsov,~S.~L.; Laubschat,~C.;
  Bauer,~A.; Kaindl,~G. Intercalation of Copper Underneath a Monolayer of Graphite on Ni(111). \emph{Phys. Rev. B} \textbf{2001}, \emph{64},
  035405\relax
\mciteBstWouldAddEndPuncttrue
\mciteSetBstMidEndSepPunct{\mcitedefaultmidpunct}
{\mcitedefaultendpunct}{\mcitedefaultseppunct}\relax
\EndOfBibitem
\bibitem[Varykhalov \latin{et~al.}(2008)Varykhalov, Sanchez-Barriga, Shikin,
  Biswas, Vescovo, Rybkin, Marchenko, and Rader]{Varykhalov:2008}
Varykhalov,~A.; Sanchez-Barriga,~J.; Shikin,~A.~M.; Biswas,~C.; Vescovo,~E.;
  Rybkin,~A.; Marchenko,~D.; Rader,~O. Electronic and Magnetic Properties of Quasifreestanding Graphene on Ni. \emph{Phys. Rev. Lett.} \textbf{2008},
  \emph{101}, 157601\relax
\mciteBstWouldAddEndPuncttrue
\mciteSetBstMidEndSepPunct{\mcitedefaultmidpunct}
{\mcitedefaultendpunct}{\mcitedefaultseppunct}\relax
\EndOfBibitem
\bibitem[Varykhalov \latin{et~al.}(2010)Varykhalov, Scholz, Kim, and
  Rader]{Varykhalov:2010a}
Varykhalov,~A.; Scholz,~M.; Kim,~T.; Rader,~O. Effect of Noble-Metal Contacts on Doping and Band Gap of Graphene. \emph{Phys. Rev. B}
  \textbf{2010}, \emph{82}, 121101\relax
\mciteBstWouldAddEndPuncttrue
\mciteSetBstMidEndSepPunct{\mcitedefaultmidpunct}
{\mcitedefaultendpunct}{\mcitedefaultseppunct}\relax
\EndOfBibitem
\bibitem[Voloshina \latin{et~al.}(2011)Voloshina, Generalov, Weser,
  B{\"o}ttcher, Horn, and Dedkov]{Voloshina:2011NJP}
Voloshina,~E.~N.; Generalov,~A.; Weser,~M.; B{\"o}ttcher,~S.; Horn,~K.;
  Dedkov,~Y.~S. Structural and Electronic Properties of the Graphene/Al/Ni(111) Intercalation System. \emph{New J. Phys.} \textbf{2011}, \emph{13}, 113028\relax
\mciteBstWouldAddEndPuncttrue
\mciteSetBstMidEndSepPunct{\mcitedefaultmidpunct}
{\mcitedefaultendpunct}{\mcitedefaultseppunct}\relax
\EndOfBibitem
\bibitem[Omiciuolo \latin{et~al.}(2014)Omiciuolo, ndez, Miniussi, Orlando,
  Lacovig, Lizzit, scedil, Locatelli, Larciprete, Bianchi, Ulstrup, Hofmann,
  egrave, and Baraldi]{Omiciuolo:2014dn}
Omiciuolo,~L.; ndez,~E. R. H.~a.; Miniussi,~E.; Orlando,~F.; Lacovig,~P.;
  Lizzit,~S.; scedil,~T. O.~M.; Locatelli,~A.; Larciprete,~R.; Bianchi,~M.;
  Ulstrup,~S. o.~r.; Hofmann,~P.; egrave,~D.~A.; Baraldi,~A. Bottom-Up Approach for the Low-Cost Synthesis of Graphene-Alumina Nanosheet Interfaces Using Bimetallic Alloys. \emph{Nat.
  Commun.} \textbf{2014}, \emph{5}, 5062\relax
\mciteBstWouldAddEndPuncttrue
\mciteSetBstMidEndSepPunct{\mcitedefaultmidpunct}
{\mcitedefaultendpunct}{\mcitedefaultseppunct}\relax
\EndOfBibitem
\bibitem[Dedkov \latin{et~al.}(2017)Dedkov, Klesse, Becker, Sp{\"a}th, Papp,
  and Voloshina]{Dedkov:2017jn}
Dedkov,~Y.; Klesse,~W.; Becker,~A.; Sp{\"a}th,~F.; Papp,~C.; Voloshina,~E. Decoupling of Graphene from Ni(111) via Formation of an Interfacial NiO Layer.
  \emph{Carbon} \textbf{2017}, \emph{121}, 10--16\relax
\mciteBstWouldAddEndPuncttrue
\mciteSetBstMidEndSepPunct{\mcitedefaultmidpunct}
{\mcitedefaultendpunct}{\mcitedefaultseppunct}\relax
\EndOfBibitem
\bibitem[Larciprete \latin{et~al.}(2016)Larciprete, Colonna, Ronci, Flammini,
  Lacovig, Apostol, Politano, Feulner, Menzel, and Lizzit]{Larciprete:2016gf}
Larciprete,~R.; Colonna,~S.; Ronci,~F.; Flammini,~R.; Lacovig,~P.; Apostol,~N.;
  Politano,~A.; Feulner,~P.; Menzel,~D.; Lizzit,~S. Self-Assembly of Graphene Nanoblisters Sealed to a Bare Metal
Surface. \emph{Nano Lett.}
  \textbf{2016}, \emph{16}, 1808--1817\relax
\mciteBstWouldAddEndPuncttrue
\mciteSetBstMidEndSepPunct{\mcitedefaultmidpunct}
{\mcitedefaultendpunct}{\mcitedefaultseppunct}\relax
\EndOfBibitem
\bibitem[Bignardi \latin{et~al.}(2017)Bignardi, Lacovig, Dalmiglio, Orlando,
  Ghafari, Petaccia, Baraldi, Larciprete, and Lizzit]{Bignardi:2017aa}
Bignardi,~L.; Lacovig,~P.; Dalmiglio,~M.~M.; Orlando,~F.; Ghafari,~A.;
  Petaccia,~L.; Baraldi,~A.; Larciprete,~R.; Lizzit,~S. Key Role of Rotated Domains in Oxygen Intercalation at Graphene on Ni(111). \emph{2D Materials}
  \textbf{2017}, \emph{4}, 025106\relax
\mciteBstWouldAddEndPuncttrue
\mciteSetBstMidEndSepPunct{\mcitedefaultmidpunct}
{\mcitedefaultendpunct}{\mcitedefaultseppunct}\relax
\EndOfBibitem
\bibitem{Schroder:2016eb}
Schr\"oder,~U.~A.; Petrovic,~M.; Gerber, T.; Martinez-Galera,~A.~J.; Gr\aa n\"as,~E.; Arman,~M.~A.; Herbig,~C.; Schnadt,~J.; Kralj,~M.; Knudsen,~J; Michely,~T. Core Level Shifts of Intercalated Graphene. \emph{2D Materials}
  \textbf{2017}, \emph{4}, 015013\relax
\mciteBstWouldAddEndPuncttrue
\mciteSetBstMidEndSepPunct{\mcitedefaultmidpunct}
{\mcitedefaultendpunct}{\mcitedefaultseppunct}\relax
\EndOfBibitem
\bibitem[D{\"u}rr \latin{et~al.}(1997)D{\"u}rr, van~der Laan, Spanke,
  Hillebrecht, and Brookes]{Durr:1997fa}
D{\"u}rr,~H.~A.; van~der Laan,~G.; Spanke,~D.; Hillebrecht,~F.~U.;
  Brookes,~N.~B. Electron-Correlation-Induced Magnetic Order of Ultrathin Mn Films. \emph{Phys. Rev. B} \textbf{1997}, \emph{56}, 8156--8162\relax
\mciteBstWouldAddEndPuncttrue
\mciteSetBstMidEndSepPunct{\mcitedefaultmidpunct}
{\mcitedefaultendpunct}{\mcitedefaultseppunct}\relax
\EndOfBibitem
\bibitem[Allen and Venus(2001)Allen, and Venus]{Allen:2001}
Allen,~M.; Venus,~D. Ordered Alloy Films of Ni and Mn Grown on Ni(111)/W(110). \emph{Surf. Sci.} \textbf{2001}, \emph{477},
  209--218\relax
\mciteBstWouldAddEndPuncttrue
\mciteSetBstMidEndSepPunct{\mcitedefaultmidpunct}
{\mcitedefaultendpunct}{\mcitedefaultseppunct}\relax
\EndOfBibitem
\bibitem[Rader \latin{et~al.}(2001)Rader, Mizokawa, Fujimori, and
  Kimura]{Rader:2001di}
Rader,~O.; Mizokawa,~T.; Fujimori,~A.; Kimura,~A. Structure and Electron Correlation of Mn on Ni(110). \emph{Phys. Rev. B}
  \textbf{2001}, \emph{64}, 165414\relax
\mciteBstWouldAddEndPuncttrue
\mciteSetBstMidEndSepPunct{\mcitedefaultmidpunct}
{\mcitedefaultendpunct}{\mcitedefaultseppunct}\relax
\EndOfBibitem
\bibitem[Manju \latin{et~al.}(2010)Manju, Topwal, Rossi, and
  Vobornik]{Manju:2010cs}
Manju,~U.; Topwal,~D.; Rossi,~G.; Vobornik,~I. Electronic Structure of the Two-Dimensionally Ordered Mn/Cu(110) Magnetic Surface Alloy. \emph{Phys. Rev. B}
  \textbf{2010}, \emph{82}, 035442\relax
\mciteBstWouldAddEndPuncttrue
\mciteSetBstMidEndSepPunct{\mcitedefaultmidpunct}
{\mcitedefaultendpunct}{\mcitedefaultseppunct}\relax
\EndOfBibitem
\bibitem[Voloshina \latin{et~al.}(2013)Voloshina, Ovcharenko, Shulakov, and
  Dedkov]{Voloshina:2013cw}
Voloshina,~E.; Ovcharenko,~R.; Shulakov,~A.; Dedkov,~Y. Theoretical Description of X-ray Absorption Spectroscopy
of the Graphene-Metal Interfaces. \emph{J. Chem. Phys.}
  \textbf{2013}, \emph{138}, 154706\relax
\mciteBstWouldAddEndPuncttrue
\mciteSetBstMidEndSepPunct{\mcitedefaultmidpunct}
{\mcitedefaultendpunct}{\mcitedefaultseppunct}\relax
\EndOfBibitem
\bibitem[Matsumoto \latin{et~al.}(2013)Matsumoto, Entani, Koide, Ohtomo,
  Avramov, Naramoto, Amemiya, Fujikawa, and Sakai]{Matsumoto:2013eu}
Matsumoto,~Y.; Entani,~S.; Koide,~A.; Ohtomo,~M.; Avramov,~P.~V.; Naramoto,~H.;
  Amemiya,~K.; Fujikawa,~T.; Sakai,~S. Spin Orientation Transition Across the Single-Layer Graphene/Nickel Thin Film Interface. \emph{J. Mater. Chem. C} \textbf{2013},
  \emph{1}, 5533--5537\relax
\mciteBstWouldAddEndPuncttrue
\mciteSetBstMidEndSepPunct{\mcitedefaultmidpunct}
{\mcitedefaultendpunct}{\mcitedefaultseppunct}\relax
\EndOfBibitem
\bibitem[Verbitskiy \latin{et~al.}(2015)Verbitskiy, Fedorov, Profeta, Stroppa,
  Petaccia, Senkovskiy, Nefedov, W{\"o}ll, Usachov, Vyalikh, Yashina, Eliseev,
  Pichler, and Gr{\"u}neis]{Verbitskiy:2015kq}
Verbitskiy,~N.~I.; Fedorov,~A.~V.; Profeta,~G.; Stroppa,~A.; Petaccia,~L.;
  Senkovskiy,~B.; Nefedov,~A.; W{\"o}ll,~C.; Usachov,~D.~Y.; Vyalikh,~D.~V.;
  Yashina,~L.~V.; Eliseev,~A.~A.; Pichler,~T.; Gr{\"u}neis,~A. Atomically Precise
Semiconductor-Graphene and hBN
Interfaces by Ge Intercalation. \emph{Sci. Rep.}
  \textbf{2015}, \emph{5}, 17700\relax
\mciteBstWouldAddEndPuncttrue
\mciteSetBstMidEndSepPunct{\mcitedefaultmidpunct}
{\mcitedefaultendpunct}{\mcitedefaultseppunct}\relax
\EndOfBibitem
\bibitem[Bertoni \latin{et~al.}(2004)Bertoni, Calmels, Altibelli, and
  Serin]{Bertoni:2004}
Bertoni,~G.; Calmels,~L.; Altibelli,~A.; Serin,~V. First-Principles Calculation of the Electronic Structure and EELS Spectra
at the Graphene/Ni(111) Interface. \emph{Phys. Rev. B}
  \textbf{2004}, \emph{71}, 075402\relax
\mciteBstWouldAddEndPuncttrue
\mciteSetBstMidEndSepPunct{\mcitedefaultmidpunct}
{\mcitedefaultendpunct}{\mcitedefaultseppunct}\relax
\EndOfBibitem
\bibitem[Achilli \latin{et~al.}(2018)Achilli, Tognolini, Fava, Ponzoni, Drera,
  Cepek, Patera, Africh, Castillo, Trioni, and Pagliara]{Achilli:2018if}
Achilli,~S.; Tognolini,~S.; Fava,~E.; Ponzoni,~S.; Drera,~G.; Cepek,~C.;
  Patera,~L.~L.; Africh,~C.; Castillo,~E.~d.; Trioni,~M.~I.; Pagliara,~S.
  Surface States Characterization in the Strongly Interacting Graphene/Ni(111) System. \emph{New J. Phys.} \textbf{2018}, \emph{20}, 103039\relax
\mciteBstWouldAddEndPuncttrue
\mciteSetBstMidEndSepPunct{\mcitedefaultmidpunct}
{\mcitedefaultendpunct}{\mcitedefaultseppunct}\relax
\EndOfBibitem
\bibitem[Kreutz \latin{et~al.}(1998)Kreutz, Greber, Aebi, and
  Osterwalder]{Kreutz:1998aa}
Kreutz,~T.~J.; Greber,~T.; Aebi,~P.; Osterwalder,~J. Temperature-Dependent Electronic Structure of Nickel Metal. \emph{Phys. Rev. B} \textbf{1998}, \emph{58},
  1300--1317\relax
\mciteBstWouldAddEndPuncttrue
\mciteSetBstMidEndSepPunct{\mcitedefaultmidpunct}
{\mcitedefaultendpunct}{\mcitedefaultseppunct}\relax
\EndOfBibitem
\bibitem[Mulazzi \latin{et~al.}(2006)Mulazzi, Hochstrasser, Corso, Vobornik,
  Fujii, Osterwalder, Henk, and Rossi]{Mulazzi:2006fw}
Mulazzi,~M.; Hochstrasser,~M.; Corso,~M.; Vobornik,~I.; Fujii,~J.;
  Osterwalder,~J.; Henk,~J.; Rossi,~G. Matrix Element Effects in Angle-Resolved Valence Band Photoemission with Polarized Light
from the Ni(111) Surface. \emph{Phys. Rev. B} \textbf{2006},
  \emph{74}, 035118\relax
\mciteBstWouldAddEndPuncttrue
\mciteSetBstMidEndSepPunct{\mcitedefaultmidpunct}
{\mcitedefaultendpunct}{\mcitedefaultseppunct}\relax
\EndOfBibitem
\bibitem[Kulkova \latin{et~al.}(2002)Kulkova, Valujsky, Kim, Condensed, and
  {2002}]{Kulkova:2002gn}
Kulkova,~S.~E.; Valujsky,~D.~V.; Kim,~J.-S.; Lee, G.; Koo, Y. M. The Electronic Properties of FeCo, Ni$_3$Mn and Ni$_3$Fe at the Order-Disorder Transition. \emph{Physica B},
  \textbf{2002}, \emph{322}, 236--247\relax
\mciteBstWouldAddEndPuncttrue
\mciteSetBstMidEndSepPunct{\mcitedefaultmidpunct}
{\mcitedefaultendpunct}{\mcitedefaultseppunct}\relax
\EndOfBibitem
\bibitem[Palumbo \latin{et~al.}(2014)Palumbo, Fries, Corso, K{\"o}rmann,
  Hickel, and Neugebauer]{Palumbo:2014ks}
Palumbo,~M.; Fries,~S.~G.; Corso,~A.~D.; K{\"o}rmann,~F.; Hickel,~T.;
  Neugebauer,~J. Reliability Evaluation of Thermophysical
Properties from First-Principles Calculations. \emph{J. Phys.: Condens. Matter} \textbf{2014}, \emph{26},
  335401\relax
\mciteBstWouldAddEndPuncttrue
\mciteSetBstMidEndSepPunct{\mcitedefaultmidpunct}
{\mcitedefaultendpunct}{\mcitedefaultseppunct}\relax
\EndOfBibitem
\bibitem[Tian \latin{et~al.}(1992)Tian, Li, Wu, Jona, and Marcus]{Tian:1992aa}
Tian,~D.; Li,~H.; Wu,~S.~C.; Jona,~F.; Marcus,~P.~M. Atomic and Alectronic Structure of Thin Films of Mn on Pd\{111\}. \emph{Phys. Rev. B}
  \textbf{1992}, \emph{45}, 3749--3754\relax
\mciteBstWouldAddEndPuncttrue
\mciteSetBstMidEndSepPunct{\mcitedefaultmidpunct}
{\mcitedefaultendpunct}{\mcitedefaultseppunct}\relax
\EndOfBibitem
\bibitem[Kresse and Hafner(1994)Kresse, and Hafner]{Kresse:1994}
Kresse,~G.; Hafner,~J. Norm-Conserving and Ultrasoft Pseudopotentials for First-Row and Transition Elements. \emph{J. Phys.: Condens. Matter} \textbf{1994},
  \emph{6}, 8245--8257\relax
\mciteBstWouldAddEndPuncttrue
\mciteSetBstMidEndSepPunct{\mcitedefaultmidpunct}
{\mcitedefaultendpunct}{\mcitedefaultseppunct}\relax
\EndOfBibitem
\bibitem[Kresse and Furthmuller(1996)Kresse, and Furthmuller]{Kresse:1996a}
Kresse,~G.; Furthmuller,~J. Efficiency of Ab-Initio Total Energy Calculations for Metals and Semiconductors Using a Plane-Wave Basis Set. \emph{Comp. Mater. Sci.} \textbf{1996}, \emph{6},
  15--50\relax
\mciteBstWouldAddEndPuncttrue
\mciteSetBstMidEndSepPunct{\mcitedefaultmidpunct}
{\mcitedefaultendpunct}{\mcitedefaultseppunct}\relax
\EndOfBibitem
\bibitem[Perdew \latin{et~al.}(1996)Perdew, Burke, and Ernzerhof]{Perdew:1996}
Perdew,~J.; Burke,~K.; Ernzerhof,~M. Generalized Gradient Approximation Made Simple. \emph{Phys. Rev. Lett.} \textbf{1996},
  \emph{77}, 3865--3868\relax
\mciteBstWouldAddEndPuncttrue
\mciteSetBstMidEndSepPunct{\mcitedefaultmidpunct}
{\mcitedefaultendpunct}{\mcitedefaultseppunct}\relax
\EndOfBibitem
\bibitem[Bl{\"o}chl(1994)]{Blochl:1994}
Bl{\"o}chl,~P.~E. Projector Augmented-Wave Method. \emph{Phys. Rev. B} \textbf{1994}, \emph{50},
  17953--17979\relax
\mciteBstWouldAddEndPuncttrue
\mciteSetBstMidEndSepPunct{\mcitedefaultmidpunct}
{\mcitedefaultendpunct}{\mcitedefaultseppunct}\relax
\EndOfBibitem
\bibitem[Bl{\"o}chl \latin{et~al.}(1994)Bl{\"o}chl, Jepsen, and
  Andersen]{Blochl:1994vg}
Bl{\"o}chl,~P.~E.; Jepsen,~O.; Andersen,~O. Improved Tetrahedron Method for Brillouin-Zone Integrations. \emph{Phys. Rev. B} \textbf{1994},
  \emph{49}, 16223--16233\relax
\mciteBstWouldAddEndPuncttrue
\mciteSetBstMidEndSepPunct{\mcitedefaultmidpunct}
{\mcitedefaultendpunct}{\mcitedefaultseppunct}\relax
\EndOfBibitem
\bibitem[Grimme(2006)]{Grimme:2006}
Grimme,~S. Semiempirical GGA-Type Density Functional Constructed with a Long-Range Dispersion Correction. \emph{J. Comput. Chem.} \textbf{2006}, \emph{27}, 1787--1799\relax
\mciteBstWouldAddEndPuncttrue
\mciteSetBstMidEndSepPunct{\mcitedefaultmidpunct}
{\mcitedefaultendpunct}{\mcitedefaultseppunct}\relax
\EndOfBibitem
\bibitem[Medeiros \latin{et~al.}(2014)Medeiros, Stafstr{\"o}m, and
  Bj{\"o}rk]{Medeiros:2014ka}
Medeiros,~P. V.~C.; Stafstr{\"o}m,~S.; Bj{\"o}rk,~J. Effects of Extrinsic and Intrinsic Perturbations on the Electronic Structure of Graphene:
Retaining an Effective Primitive Cell Band Structure by Band Unfolding. \emph{Phys. Rev. B}
  \textbf{2014}, \emph{89}, 041407\relax
\mciteBstWouldAddEndPuncttrue
\mciteSetBstMidEndSepPunct{\mcitedefaultmidpunct}
{\mcitedefaultendpunct}{\mcitedefaultseppunct}\relax
\EndOfBibitem
\bibitem[Medeiros \latin{et~al.}(2015)Medeiros, Tsirkin, Stafstr{\"o}m, and
  Bj{\"o}rk]{Medeiros:2015ks}
Medeiros,~P. V.~C.; Tsirkin,~S.~S.; Stafstr{\"o}m,~S.; Bj{\"o}rk,~J. Unfolding Spinor Wave Functions and Expectation Values of General Operators:
Introducing the Unfolding-Density Operator.
  \emph{Phys. Rev. B} \textbf{2015}, \emph{91}, 041116--5\relax
\mciteBstWouldAddEndPuncttrue
\mciteSetBstMidEndSepPunct{\mcitedefaultmidpunct}
{\mcitedefaultendpunct}{\mcitedefaultseppunct}\relax
\EndOfBibitem
\end{mcitethebibliography}
\end{document}